\documentstyle[preprint,aps]{revtex}
\begin{document}
\preprint{OITS-551}
\draft
\title{ CP VIOLATION IN A Multi-Higgs Doublet Model\footnote{Talk presented by
Deshpande at the Conference WHEPP-3, December 1993.}}
\author{N.G. Deshpande and Xiao-Gang He}
\address{Institute of Theoretical Science\\
University of Oregon\\
Eugene, OR 97403-5203, USA}
\date{August, 1994}
\maketitle
\begin{abstract}
We study CP violation in a multi-Higgs doublet model
based on a $S_3 \times Z_3$ horizontal symmetry. We consider two mechanisms
for CP violation in this model: a) CP violation due to complex Yukawa
couplings;
and b) CP violation due to scalar-pseudoscalar mixings.  We find that the
predictions for
$\epsilon'/\epsilon$, CP
violation in B decays and the electric dipole moments of neutron and
electron are different between these two mechanisms.
These predictions are also dramatically different from the minimal Standard
Model predictions.
\end{abstract}
\pacs{}
\newpage

\section{Introduction}

The origin of CP violation, is one of the
outstanding problems of particle physics today. So far CP
violation has only been observed in the neutral Kaon system. The observed CP
violation can be explained in many models. It is therefore important to study
other CP violating experimental observables and compare the results with
different model predictions.  Such study may reveal the real origin of CP
violation.

In the minimal $SU(3)_C\times SU(2)_L\times U(1)_Y$  Standard Model (MSM),
there is only one Higgs doublet. When the Higgs develops valcuum expectation
value (VEV) v, all fermion receive masses. In the mass eigenstate basis, the
Higgs coupling to fermions is diagonal, it does not mediate CP violating
interaction. However, the coupling of
the charged current to the W-boson becomes complex. It is given by
\begin{eqnarray}
L_C = {g\over \sqrt{2}} \bar u_i V_{ij}\gamma^\mu {1-\gamma_5\over 2}
 d_j W^+_\mu + H.C.\;,
\end{eqnarray}
where the matrix $V_{ij}$ is the CKM matrix $V_{KM}$\cite{km}. For three
generations of
quarks, there is a non-removable phase in the matrix. This is the source of
CP violation in the MSM. This matrix is conveniently parametrized as, following
Wolfenstein \cite{wolf}
\begin{eqnarray}
V_{KM} = \left ( \begin{array}{ccc}
1-\lambda^2/2& \lambda &A\lambda^3(\rho - i \eta)\\
-\lambda&1-\lambda^2/2-iA^2\lambda^4\eta& A\lambda^2\\
A\lambda^3(1-\rho-i\eta)& -A\lambda^2& 1
\end{array} \right )\;,
\end{eqnarray}
where $\lambda = V_{us} = 0.221$. If $\eta \neq 0$, CP is violated. Unitarity
constraints on the matrix elements provide very powerful and interesting
realtions. The most interesting one is the triangle defined by
\begin{eqnarray}
V_{ud}V^*_{td} + V_{us}V^*_{ts} + V_{ub}V^*_{tb} = 0\;.
\end{eqnarray}
In the Wolfenstein parametrization, $V_{ud} \approx V_{tb} \approx 1$ and
$V_{ts}\approx V^*_{cb}$, we have
\begin{eqnarray}
V_{ub} +V^*_{td} \approx V_{ub}V_{cb}\;.
\end{eqnarray}
This defines the triangle shown in Figure 1 with three angles
$\alpha$, $\beta$ and $\gamma$. The area of the triangle is given
by $A^2\lambda^6\eta/2$. CP violation in the neutral Kaon system can be
explained by the "box" interaction\cite{lim}. If CP violation due to the phase
in the CKM matrix is the only source for CP violation,
experiments at B factories will be able to determine all the three
angles\cite{bcp}.

Another class of model for CP violation is the multi-Higgs doublet model. If
there are more than one Higgs doublets, the neutral scalar couplings to the
quarks are not necessarily diagonal, and therefore provide new sources for CP
violation\cite{fcp}.
CP violation can arise in three places in
this type of models: 1) Non-trivial phase in the $V_{KM}$ matrix; 2)Non-trivial
phases in the Yukawa couplings; and 3) Mixings of scalar and
pseudoscalar Higgs bosons. In cases 2) and
3), CP violation can occurs at the tree level by exchanging neutral Higgs
bosons.
 In this talk we will present some studies of CP violation in multi-Higgs
models with flavour changing neutral currents at the trre level which has CP
violation predominantly throug mechanisms 2) and 3).

A most general study suffers from too many free parameters. To have a definite
idea, we
carry out the study in a $S_3\times Z_3$ model proposed by Ma\cite{ma}. This
model  has some very interesting predictions for fermion masses and their
mixings. It  also has interesting predictions for CP
violation\cite{desh,lav,hdesh}. We study the predictions for: (i)
$\epsilon'/\epsilon$; (ii) CP violation in the neutral B
system; and (iii) the neutron and electron electric dipole moments (EDM). We
compare these predictions with those in the MSM.

\section
{Yukawa couplings in the $S_3\times Z_3$ model}

In the $S_3\times Z_3$ model, there are four Higgs doublets, $\phi_{1,2,3,4}$.
The quarks and Higgs bosons transform under the $S_3\times Z_3$ symmetry
as\cite{ma}
\begin{eqnarray}
q_{3L}\;, t_R\;, b_R\;, \phi_1: (1, 1)\;\;\;\;\;&,&(q_{1L}, q_{2L})\;, (\phi_3,
\phi_4):  (2, \omega)\;,\nonumber\\
 (c_R, u_R)\;, (s_R, d_R): (2,\omega^2)\;&,&\phi_2: (1,\omega^2)\;,
\end{eqnarray}
where $\omega \neq 1$, $\omega^3 = 1$ is the $Z_3$ element.
The Yukawa couplings consistent with the $S_3\times Z_3$ symmetry are given by
\begin{eqnarray}
L_Y &=& -f_1(\bar q_{1L} \tilde \phi_3 u_R + \bar q_{2L} \tilde \phi_4 c_R)
-f_2 \bar q_{3L}\tilde \phi_1 t_R - f_3 (\bar q_{1L}\phi_2 s_R + \bar q_{2L}
\phi_2 d_R)\nonumber\\
&-&f_4(\bar q_{1L}\phi_3 b_R + \bar q_{2L} \phi_4 b_R)
-f_5(\bar q_{3L}\phi_3 d_R + \bar q_{3L} \phi_4 s_R) -f_6 \bar q_{3L}\phi_1
b_R
+ H.C.
\end{eqnarray}
where $\tilde \phi_i = (\phi^{0*}_i, -\phi_i^-)^T$. Without loss of generality
we work in a basis where all VEVs are real. The up-quark mass matrix is
diagonal : $\hat M^u = Diag(f_1v_3, f_1v_4, f_2v_1)$, and the down-quark mass
matrix can be written as, with a suitable choice of quark phases,
\begin{eqnarray}
M^d = \left ( \matrix{ 0&a&\xi b\cr a&0&b\cr \xi c&c&d\cr}\right )\;,
\end{eqnarray}
with a, b, c, d real and $\xi = |\xi|e^{i\sigma}$ complex. $M^d$ can be
diagonalized by a bi-unitary transformation
$M^d = V_L \hat M^d V^\dagger_R$.
Here $\hat M^d$ is the diagonalized down quark mass matrix. $V_L$ and $V_R$ are
unitary matrices. Because the up quark mass matrix is already diagonalized,
$V_L
$ is the CKM matrix $V_{KM}$.

 It is convenient to work in a basis of the Higgs bosons in which the Goldstone
bosons are removed. To this end we define the
following\cite{lav}
\begin{eqnarray}
\left ( \matrix{\phi_1\cr \phi_2\cr \phi_3\cr \phi_4\cr} \right )
= \left ( \matrix{ {v_1\over v} & {v_2\over v_{12}} & {v_1v_4\over v_{12}
v_{124}} & -{v_1v_3\over vv_{124}}\cr
{v_2\over v} & - {v_1\over v_{12}} & {v_2v_4\over v_{12}v_{124}} &
-{v_2v_3\over v v_{124}}\cr
{v_3\over v} &0&0& {v_{124}\over v}\cr
{v_4\over v} &0& -{v_{12}\over v_{124}}& - {v_3v_4\over v v_{124}}\cr} \right )
\left (\matrix {G\cr H_1\cr H_2\cr H_3\cr} \right )\;,
\end{eqnarray}
where $v_{12}^2 = v_1^2+v_2^2$, $v_{124}^2 = v_1^2 +v_2^2+v_4^2$, and $v^2 =
v_1^2 +v_2^2+v_3^2+v_4^2$.
The transformation is the same for both the neutral and charged Higgs bosons.
For the
neutral Higgs bosons, $G = h^0 + iG_Z$, where $G_Z$ is the Goldstone boson
'eaten'
by Z, and $h^0$ is a physical field whose couplings are the same as the Higgs
boson in the MSM. For the charged Higgs bosons G is the
Goldstone boson 'eaten' by W. In this basis, we have

\begin{eqnarray}
L_Y &=& -(\bar D_L \hat M^d D_R + \bar U_L \hat
M^u U_R)(1 + {Reh^0\over v\sqrt{2}})
\nonumber\\
&-& \bar D_L \tilde Y^d_i D_R {h^0_i\over \sqrt{2}}
- \bar U_L \tilde Y^u_i  U_R
{h^{0*}_i\over \sqrt{2}}\nonumber\\
&-& \bar U_L V_{KM} \tilde Y^d_i D_R h^+_i +
 \bar D_L V^\dagger_{KM} \tilde Y^u_i U_R h^-_i + H.C.\;,
\end{eqnarray}
where $h_i$ are the component fields of $H_i$ with $H_i = (h^+_i,
h^0_i/\sqrt{2})$. $U_{L,R} = (u, c, t)_{L,R}^T$, and $D_{L,R} = (d, s,
b)_{L,R}^T$. The couplings $\tilde Y_i$ can be easily expressed in term of
quark masses, $V_{L,R}$, and VEVs.

In general $h_i^{0,+}$ are not the mass eigenstates. We can parametrize the
mixings as
\begin{eqnarray}
\left (\matrix{h^0\cr Reh^0_k\cr Imh^0_k\cr}\right )
&=& \left ( \matrix{\alpha_{00}&\alpha_{0i}&\beta'_{0j}\cr
\alpha_{k0}&\alpha_{ki}&\beta'_{kj}\cr
\alpha'_{k0}&\alpha'_{ki} & \beta_{kj}\cr} \right ) \left (\matrix{R_0\cr
R_i\cr I_j\cr}
\right )\;,\nonumber\\
\\
h^+_i &=& (\gamma_{ij})\eta^+_j\nonumber\;,
\end{eqnarray}
where $R_i$, $I_i$ and $\eta_i$ are the mass eigenstates, the matrix $(\alpha
\beta)$ is a $7\times7$ othogonal matrix, and $(\gamma)$ is a $3\times 3$
unitary matrix.

The specific values for the mixings depend on the details
of the Higgs potential. Unfortunately they are not determined. To simplify the
problem,
we will discuss two cases: a) CP violation only comes from complex Yukawa
couplings; and b) CP violation only comes from the mixings of real and
imaginary
$h^0_i$\cite{hdesh}. Case a) can be realised by constraining certain soft
symmetry breaking
terms in the potential\cite{lav}. We further assume, for simplicity, that
$Reh^0_i$ are the mass
eigenstate $R_i$ and consider their effects. The same analysis can be easily
carried out for $Imh^0_i$ in the same way. The source for CP violation is the
non-zero value
for $\sigma$ which is a free parameter.  We will present our
results for $\sigma = 80^0$, which is close to the maximum of the allowed
phase. Case b) can be realised by requiring spontaneous CP violation. The value
of $\sigma$ will be zero and CP violation arises due to scalar-pesudoscalar
Higgs boson mixing. For illustration, we consider the effects of a neutral
mixed state
\begin{eqnarray}
R = cos\theta Reh^0_2 + sin\theta Imh^0_3\;,
\end{eqnarray}
and for the charged Higgs boson we consider mixing
\begin{eqnarray}
\eta^+ = \gamma_{22}h^+_2 + \gamma_{23} h^+_3\;,
\end{eqnarray}
where $\gamma_{ij}$ are complex numbers, and
$|\gamma_{22}|^2 +|\gamma_{23}|^2 = 1$.

The parameters $a$, $b$, $c$, and $d$ are constrained from the down quark
masses and the CKM mixings. We take as input parameters $a =
0.04 GeV$, $b = 0.25 GeV$, $c= 2.66 GeV$ , $d = 4 GeV$.
The mass eigenvalues for the down quarks are quite insensitive to the phase
$\sigma$. For both cases, we have $m_b =4.8GeV$, $m_s = 149 MeV$ and $m_d = 9.5
MeV$. These values are well within the allowed regions\cite{qm}. The CKM matrix
for case a) is
\begin{eqnarray}
V_{KM} = \left ( \matrix{0.975&-0.222&0.00476\cr
0.221 +i0.0033&0.974+i0.014& 0.043-i0.0015\cr
-0.014+ i 1.2\times 10^{-5}&-0.041-i6.8\times 10^{-4}& 0.998 - i0.034\cr}
\right )\;,
\end{eqnarray}
and for case b)
\begin{eqnarray}
V_{KM} = \left ( \matrix{0.975&0.22&0.0048\cr
-0.219&0.975&-0.0436\cr-0.014&0.0415&0.999}\right)\;.
\end{eqnarray}

The values for the VEV's are not fixed, we only know
$v_3/v_4 = m_u/m_c$. We will use the values: $v_1 =v_2 = 44
GeV$, $v_3 = 0.9$ GeV and $v_4 = 238 GeV$ for illustration. We shall comment on
effects of changing these values later.

\section
{ Constraints on the Higgs boson masses from the  neutral K and B meson
systems}

The $S_3\times Z_3$ model has very restrictive allowed values for the
non-trivial CP violating phase in the CKM matrix. The CP violating measure
J\cite{jask} is less than $2.5 \times 10^{-6}$ which is too small to explain CP
violation in the neutral Kaon sysytem. Therefore in this model CP
violation due to Higgs boson exchange has to be
considered.

 The CP violating parameter $\bar\epsilon$
is given by
\begin{eqnarray}
\bar\epsilon = {ImM_{12}^K\over \sqrt{2}\Delta m_K}e^{i\pi/4}\;,
\end{eqnarray}
where $M_{12}^K$ is the matrix element which mixes $K^0$ with $\bar K^0$, and
$\Delta m_K$ is the mass difference between $m_{K_L}$ and $m_{K_S}$.
Experimental value for $\bar \epsilon$ is $2.3\times 10^{-3} e ^{i\pi/4}$. The
$\Delta S = 2$ Hamiltonian, responsible for $M^K_{12}$, generated by exchanging
neutral Higgs bosons $R_i$ is given by
\begin{eqnarray}
H_{eff} = -{1\over 2M^2_{R_i}}
\left ( \bar d [(\alpha_{ki} +i \alpha'_{ki})\tilde Y^d_{k,12})
{1+\gamma_5\over 2}
+ (\alpha_{ki} -i\alpha'_{ki})\tilde Y^{d*}_{k,21}){1-\gamma_5\over 2}]s\right
)^2\;.
\end{eqnarray}
We obtain
\begin{eqnarray}
M_{12}^K&=& <K^0|H_{eff}|\bar K^0>\nonumber\\
&=& -{f_k^2m_K\over 2 M^2_{R_i}}
( -{5\over 24}{m_K^2\over (m_s+m_d)^2}[(\alpha_{ki} +i \alpha'_{ki})\tilde
Y^d_{k,12})^2
+ (\alpha_{ki} -i\alpha'_{ki})\tilde Y^{d*}_{k,21})^2]
\nonumber\\
&+&(\alpha_{ki} +i\alpha'_{ki})\tilde Y^d_{k,12}(\alpha_{k'i}
-i\alpha'_{k'i})\tilde Y^{d*}_{k',21}
({1\over 12} + {1\over 2} {m_K^2\over (m_s+m_d)^2}) )
\;.
\end{eqnarray}
Here we have used the vaccum saturation and factorization approximation results
for the matrix elements\cite{tram}.
The contribution to the mass difference $\Delta m_K$ is given
by $2ReM_{12}$. Similar formula holds for the neutral B system.

To constrain the Higgs boson masses, we require that the neutral Higgs boson
contributions to the mass differences in
the neutral K and B systems to be less than the experimental values:
$\Delta m_K/m_K = 7 \times 10^{-15}$, and $\Delta m_B/m_B = 8 \times 10^{-14}$.
We find that for case a) the tightest constraints on the masses of
$Reh^0_{1,2}$ are from the mass difference $\Delta M_B$ of the neutral  B
mesons which gives $M_{h_1} >2.9 TeV$ and $ M_{h_2}>3.1 TeV$. With these
masses, $Reh^0_{1,2}$ can
not produce large enough
$\bar \epsilon$. Similar consideration yields $M_{h3} > 3.5 TeV$, and we find
the experimental value of $\bar\epsilon$ can now be
produced if the mass is about  $5.6 TeV$. The mass difference
$\Delta M_K$ of the neutral K mesons gives weaker
bounds in all cases.
For case b),  the
experimental value of $\Delta M_B$ constrains $M_R > 3 TeV$. From the
experimental value of $\bar\epsilon$, we obtain $sin\theta
cos\theta/M^2_R = 1.1\times 10^{-8}\,{\rm GeV}^{-2} $ which implies $M_R < 7
TeV$.

\section
{Predictions for $\epsilon'/\epsilon$.}

In this section we study the direct CP violation in $K_{L,S}\rightarrow 2\pi$
decays. CP violation in these processes is characterized by the value of
$\epsilon'/ \epsilon$. $\epsilon'/\epsilon$ is defined as
\begin{eqnarray}
{\epsilon'\over \epsilon} = {\omega\xi - ImA_2/ReA_0 \over \xi +
ImM_{12}/\Delta
M_K}\;,
\end{eqnarray}
where $\omega = ReA_2/ReA_0 = 1/20$, $\xi = ImA_0/ReA_0$. Here $A_0$ and $A_2$
are the $\Delta I = 1/2$, $3/2$ decay amplitudes for $K_{L,S}
\rightarrow 2 \pi$.

In the MSM, the contribution to $\epsilon'/\epsilon$ is donminantly
from the gluon penguin. However, for large top quark mass of order 200 GeV,
the electroweak penguin also contribute significantly and may even cancel
the gluon penguin contribution. There are large uncertainties from hadronic
matrix evaluation, $\Lambda^{(QCD)}_4$ dependce and errors in the CKM matrix.
The range for $\epsilon'/\epsilon$ is predicted to be
between  $10^{-4}$ to $10^{-3}$ for $\Lambda_4^{(QCD)}= 300$ MeV\cite{buras}.
This is consisten with the experimental constraints from
Fermilab, $(7.4\pm 6.0)\times 10^{-4}$ and CERN, $(23 \pm 6.5)\times
10^{-4}$\cite{eps}.

In the $S_3\times Z_3$ model there are several contributions to
$\epsilon'/\epsilon$. Due to large neutral Higgs masses, the neutral Higgs
boson
contributions to $\epsilon'/\epsilon$ are very small. However there may be
large contributions from the charged Higgs bosons. The dominant
contribution is from the charged Higgs boson mediated gluon
penguin. The relevant $\Delta S = 1$ effective Larangian is given by
\begin{eqnarray}
L_{\Delta S=1}= i\bar d \sigma^{\mu\nu}(\tilde f_1{1+\gamma_5\over 2}
+\tilde f_2{1-\gamma_5\over 2})\lambda^a
s G^a_{\mu\nu}\;,
\end{eqnarray}
where $G^a_{\mu\nu}$ is the gluon field strength, $\lambda^a$ are
the $SU(3)_C$ generators, and
\begin{eqnarray}
\tilde f_1 &=& {g_s(\mu)\over 32 \pi^2} {m_l\over M^2_{h^+_j}} ({3\over 2}- ln
{m_l^2\over M^2_{h^+_j}})
Im\{(V_{KM}\tilde Y^d_i \gamma_{ij})_{l1}(\tilde Y^{u\dagger}_k
V_{KM}\gamma_{kj})^*_{l2}\}\zeta_f\;,\nonumber\\
\tilde f_2 &=& {g_s(\mu)\over 32 \pi^2} {m_l\over M^2_{h^+_j}} ({3\over 2} - ln
{m_l^2\over M^2_{h^+_j}})Im\{(V_{KM}\tilde Y^d_i \gamma_{ij})^*_{l2}(\tilde
Y^{u\dagger}_k
V_{KM}\gamma_{kj})_{l1}\}\zeta_f\;,
\end{eqnarray}
where $\zeta_f= (\alpha_s(m_h)/\alpha_s(\mu))^{14/23}
 \approx 0.17$ is the QCD correction factor, and $l$ is summed over u, c and t.
We will use
$\alpha_s(\mu) \approx 4\pi/6$ for $\mu = 1 GeV$.
The above effective Lagrangian
will generate a non-zero value
for $ImA_0$\cite{sand}. $L_{\Delta S=1}$ also generates a non-zero
value $\bar \epsilon_{LD}$ for CP violation in
$K^0-\bar K^0$ mixing due to long distance
interactions through $K^0$ and $\pi$, $\eta$, $\eta'$ mixings\cite{dono}. One
obtains\cite{dono,hyc}
\begin{eqnarray}
{\xi\over \bar\epsilon_{LD}}&\approx& -0.196D\;,\nonumber\\
2m_K ImM^K_{12,LD} &\approx& 0.8 \times 10^{-7} (\tilde f_1 + \tilde
f_2)(GeV^3)\;,
\end{eqnarray}
where $D$ is a supression factor
of order $O(m^2_K, m^2_\pi)/\Lambda^2$. ${\xi/ \bar\epsilon_{LD}}$ is of order
-0.014 to -0.1.

We find that in both a) and b) cases, the donimant contributions are
from the top quark in the loop arising from mixing in the charged Higgs boson
couplings. For case a), we have
\begin{eqnarray}
\bar\epsilon_{LD}(h^+_i) &\approx& a_i{GeV^2\over m^2_{h^+_i}}
 {m_t\over 150GeV}ln{m_t^2\over m^2_{h^+_i}}\;,
\end{eqnarray}
with $a_1 = 18$, $a_2 = 25$ and $a_3 = -7$.

And for case b), we have
\begin{eqnarray}
\bar\epsilon_{LD} \approx -7.35\times 10^{3} Im(\gamma_{22}\gamma_{23}^*){GeV^2
\over
m^2_{\eta^+}}{m_t\over 150GeV}ln{m^2_t\over m^2_{\eta^+}}\;.
\end{eqnarray}
The contributions to $\bar\epsilon$ can be significant in both cases
 depending on the Higgs boson masses and the CP violating parameter
$Im(\gamma_{22}\gamma_{23}^*)$. We will study constraints on these parameters
in Sec.VI.  When these constraints are taken into account,
$\bar \epsilon_{LD}$ is generally constrained to be less than $3\times10^{-5}$
for case a) and $\epsilon'/\epsilon$ to be less than $3\times 10^{-5}$.
However, for
case b), $\bar \epsilon_{LD}$ can still be as large as $10^{-3}$ and
$\epsilon'/\epsilon$ can be $10^{-3}$.

\section
{CP violation in the neutral B system.}

There are many processes which can test CP violation in the
neutral B system. Some particularly interesting ones are\cite{bcp}
\begin{eqnarray}
B_d \rightarrow J/\psi K_S\;,\;B_d\rightarrow \pi^+\pi^-\;,\;
B_s\rightarrow \rho K_S\;.
\end{eqnarray}
The differences of time variation of decay rates for the above processes and
their
CP tranformed states are given by
\begin{eqnarray}
a_{fCP} &=& {\Gamma (B^0(t) \rightarrow f_{CP}) - \Gamma ( \bar B^0(t)
\rightarrow f_{CP})
\over \Gamma (B^0(t) \rightarrow f_{CP}) + \Gamma ( \bar B^0(t) \rightarrow
f_{CP})}
\nonumber\\
&=& {(1-| \lambda |^2) cos(\Delta M_B t) - 2Im \lambda sin(\Delta M_B t) \over
1 + | \lambda |^2}\;,
\end{eqnarray}
where $f_{CP}$ indicates the final states.  $\lambda$ is defined as
\begin{eqnarray}
\lambda = \left ({q\over p}\right )_B {\bar A \over A} S\;,
\end{eqnarray}
where $(q/p)_B = \sqrt{M^{B*}_{12}/M^B_{12}}$,  $A$ and $\bar A$ are the decay
amplitudes. If the final state contains
$K_S$, $S = (q/p)_K$ which has a phase of order $10^{-3}$. For other cases S is
equal to one.

 Non-zero asymmetry $a_{fCP}$ signals CP violation. If $|\lambda|$ is not equal
to one, it
indicates that CP is violated in the decay amplitudes. In the MSM $|\lambda|$
is
equal to one to a very good approximation for the above three processes.
The asymmetries are proportional to $Im\lambda$. In the MSM, the processes in
Eq.(24) measure the three angles $\alpha$,$\beta$ and $\gamma$,
\begin{eqnarray}
Im\lambda (B_d \rightarrow J/\psi K_S) = - sin 2 \beta\;,\nonumber\\
Im\lambda (B_d \rightarrow \pi^+ \pi^-) = sin 2\alpha\;,\\
Im\lambda (B_s \rightarrow \rho K_S) = -sin 2 \gamma\;,\nonumber
\end{eqnarray}

In the $S_3\times Z_3$ model, the situation is very different.
Although the CP
violating decay amplitudes $A$ and $\bar A$ are small, the phase of
$\sqrt{M^{B*}_{12}/M_{12}^B}$ in the $B-\bar B$ mixing due
to neutral Higgs boson exchange can be large.  In case a), there is CP
violation arising from the phase in Yukawa coupling of Higgs bosons, as well as
CKM matrix, but the former is much larger. The three meaurements in Eq.(24) do
not measure the  angles
$\alpha$, $\beta$ and $\gamma$ anymore. The first two
processes will
mostly measure the phases in $M^{B_d}_{12}$. We have
\begin{eqnarray}
Im\lambda (B_d \rightarrow \pi^+ \pi^-)
 &\approx& Im\lambda (B_d \rightarrow J/\psi K_S) \leq 0.42\;,\;\;from\;
Reh^0_1,\nonumber\\
Im\lambda(B_d \rightarrow \pi^+ \pi^-)
&\approx& Im\lambda(B_d \rightarrow J/\psi K_S) \leq 0.19\;,\;\;
from\;Reh^0_2\;,\\
Im\lambda(B_d\rightarrow \pi^+ \pi^-)
&\approx& Im\lambda(B_d \rightarrow J/\psi K_S)
\approx 0.19\;,\;\;form\;Reh^0_3\;.\nonumber
\end{eqnarray}

For case b), we find
\begin{eqnarray}
Im\lambda(B_d \rightarrow \pi^+ \pi^-)
\approx Im\lambda(B_d \rightarrow J/\psi K_S) \approx -0.25\;.
\end{eqnarray}

$Im\lambda$ for $B_s \rightarrow \rho K_S$ is different for
a) and b). For case a), the neutral Higgs boson contributions to the asymmetry
are small. However $Im\lambda(B_s \rightarrow \rho K_S)$ due to CP violation in
the KM-matrix can be
about 0.1. For case b), $Im\lambda(B_s \rightarrow \rho K_S)$
from neutral Higgs boson exchange is
only about 0.02.

If interpreted as in Eq.(27), we find for case a), $sin 2\alpha = -sin 2\beta$,
$sin\gamma = 0.05$, and $\alpha + \beta + \gamma \neq 180^0$. For case b),
we have, $sin 2\alpha = -sin 2\beta$, $sin\gamma = 0.01$. We again find,
$\alpha + \beta + \gamma \neq 180^0$.

\section
{ The neutron and electron electric dipole moments.}
The EDMs of neutron and electron in the MSM model
are extremely small. The neutron EDM $ D_n$ can only be generated at three
loop level. It is predicted to be less than $10^{-31}$ ecm\cite{he}. The
electron
EDM is even smaller ($ <10^{-36}$ ecm)\cite{krip}. The experimental upper bound
on the neutron EDM is
$1.2\times 10^{-25}$ ecm\cite{nedm}. For the electron the bound is about
$10^{-26}$ ecm\cite{eedm}. If future measurement will obtain an EDM larger than
the MSM model prediction, it will be an indication for new physics beyond the
MSM.

The prediction for the EDMs in the $S_3\times Z_3$ are very different from the
MSM. They may reach the experimental bounds.

At the one loop level, the neutral Higgs contributions to the neutron EDM are
small.
For case a) we find that $ D_n < 2\times 10^{-28}$ ecm
For case b), we have $D_n(d) \approx 2 \times 10^{-29} ecm$. The u quark EDM is
zero at the one loop
level.

There may be large contributions to
the neutron EDM at the two loop level from the
Weinberg operator\cite{wein} $D_n(W)$ and from the color dipole moment of gluon
due
to Bar-Zee type of diagrams\cite{bz,cdm} $D_n(BZ)$. In our model, we have
\begin{eqnarray}
D_n(W) &\approx& e \zeta_W \Lambda {1\over 64\pi^2} ImZ^i_{tt} {m_t^2\over
m^2_{h^0_i}}ln{m_t^2\over m_{h^0_i}^2}\;,\nonumber\\
D_n(BZ, q) &\approx& {m_q\over 64\pi^3} {c_q\over 9}
\alpha_s(\mu)\zeta_{bz}{m_t^2\over m_{h^0_i}^2}\left ( ln{m_t^2\over
m_{h^0_i}^2}\right )^2
ImZ^i_{tq}\;,
\end{eqnarray}
where $\zeta_W \approx 6\times 10^{-6}$, and $\zeta_{bz} \approx 10^{-2}$
are the QCD
correction factors, $c_u = 2$ and $c_d = 4$, and $\Lambda \approx 1 GeV$ is the
chiral symmetry breaking scale. The parameters $ImZ$ are given by
\begin{eqnarray}
ImZ_{tt}^i &=&{1\over m_t^2} Re(\tilde Y^u_{k, 33} (\alpha_{ki} - i
\alpha'_{ki})) Im(\tilde Y^u_{k',33}(\alpha_{k'i} - i
\alpha'_{k'i}))\;,\nonumber\\
ImZ^i_{tu} &=& {1\over m_u m_t} Im(\tilde Y^u_{k,33} (\alpha_{ki}
-i\alpha'_{ki})
\tilde Y^u_{k',11} (\alpha_{k'i} -i\alpha'_{k'i}))\;\nonumber\\
ImZ^i_{td} &=& {1\over m_d m_t} Im(\tilde Y^u_{k,33} (\alpha_{ki}
-i\alpha'_{ki})
\tilde Y^d_{k',11} (\alpha_{k'i} +i\alpha'_{k'i}))\;.
\end{eqnarray}

For case a), because there is no CP violation in the
up quark sector only down quark loops contribute, $D_n(W)$ from the Weinberg
operator at the
two loop
level is small.
There are non-zero $D_n(BZ)$ from d-quark due to Bar-Zee
mechanism. We
find that the contributions from $Reh^0_{1,2}$ is also small ($< 4\times
10^{-28}
ecm$). $Reh^0_3$ contribution is even smaller.

For case b), the two loop
contributions to the EDM are significantly larger because in this case there is
CP violation in the top quark interaction. We have
\begin{eqnarray}
D_n(BZ, u) \approx(2 \sim 8)\times 10^{-26} ecm\;,\nonumber\\
D_n(BZ, d) \approx (2 \sim 8)\times 10^{-27} ecm\;,
\end{eqnarray}
for $m_t$ between 100 GeV to 200 GeV.
The contribution from the Weinberg operator is small, $D_n(W) \leq 10^{-30}
ecm$.

The charged Higgs bosons can also contribute to the neutron EDM. At the one
loop level, the u  and d quark EDM are given by
\begin{eqnarray}
d_u &=& -{1\over 48\pi^2} {m_l \over m^2_{h^+_i}} ln {m^2_l\over m^2_{h^+_i}}Im
[\gamma_{ji}\gamma_{ki}^*(V_{KM}\tilde Y^d_j)_{1l} (V_{KM}^\dagger \tilde
Y^u_k)_{l1}]\;,\nonumber\\
d_d &=& {1\over 24\pi^2}{m_l \over m^2_{h^+_i}} ln {m_l^2\over m^2_{h^+_i}}
Im[\gamma_{ji}\gamma_{ki}^*(V_{KM}\tilde Y^d_j)_{l1} (V_{KM}^\dagger \tilde
Y^u_k)_{1l}]\;.
\end{eqnarray}
For $d_u$, $l$ is summed over d, s, and b; and for $d_d$, $l$ is summed over u,
c,
and t.
At the two loop level, there is a large contribution from the Weinberg
operator,
\begin{eqnarray}
D_n(W) \approx e \zeta_W' \Lambda {1\over 32\pi^2} ImZ'^i_{tt}{m_t^2\over
m^2_{h^+_i}}
ln{m_t^2\over m^2_{h^+_i}}\;,
\end{eqnarray}
where $\zeta'_W = 3 \times 10^{-4}$ is the QCD correction factor, and
\begin{eqnarray}
ImZ'^i_{tt} = {1\over m_bm_t} Im [ \gamma_{ji}\gamma_{ki}^*(V_{KM} \tilde
Y^d_j)_{33} (V_{KM}\tilde Y^u_k)_{33}]\;.
\end{eqnarray}

We find that in case a) the
dominant contributions are from the two loop Weinberg operator. We
have
\begin{eqnarray}
D_n(W) &\approx &b_i\times 10^{-19} {GeV^2\over m_{h_i^+}^2} ln {m_t^2\over
m_{h_i^+}^2}{m_t^2\over (150 GeV)^2}\;ecm\;,
\end{eqnarray}
where $b_1 = 1.6$, $b_2 = 1.4$ and $b_3 = 1.2\times 10^{-6}$.

Requiring the contributions to be less than the experimental value, we find
the masses of $h^+_{1,2}$ have to be larger than $2.5 TeV$. There is no
constraint on $h^+_3$ mass. Combining this information with those from Eqs.(22)
and (23),
we find the charged Higgs boson contributions to $\bar \epsilon_{LD}$ is less
than
$3\times 10^{-5}$, and $\epsilon'/ \epsilon$ is
less than $3 \times 10^{-5}$.

For case b), we find the dominant contribution is from the one loop d quark
EDM.
We have
\begin{eqnarray}
D_n(d) \approx 5.4 \times 10^{-19} Im(\gamma_{22}\gamma_{23}^*){GeV^2\over
m^2_{\eta^+}}
ln{m_t^2\over m_{\eta^+}^2}{m_t\over 150GeV}\; ecm\;.
\end{eqnarray}
Requiring $D_n(d)$ to be less than the experimental value,
$\bar \epsilon_{LD}$ is constrained to be less than $10^{-3}$, and $\epsilon'/
\epsilon$ can still be of order $10^{-3}$. Assuming maximum mixing, the mass
of $\eta^+$ is constrained to be larger than $5\; TeV$.

The $S_3\times Z_3$ model may also have interesting CP violating signatures in
the lepton sector.  We assume the same $S_3\times Z_3$
assignments for the left handed and the charged right handed leptons as their
quark partnenrs\cite{ma}. The mass matrix and Yukawa couplings for the charged
leptons
are similar to the down quarks. One simply changes the  parameters (a, b, c, d
, and $\xi$) for quarks to ($a_l\;, b_l\;, c_l\;, d_l\;,$ and $\xi_l =
|\xi|e^{i\sigma'}$) for leptons. We use\cite{desh}: $a_l=0.106 GeV\;, b_l=0\;,
c_l=1.781
GeV\;, d_l = 8.6\times 10^{-3} GeV$. For this set of parameters, we have $m_e =
0.511 MeV$, $m_\mu = 106 MeV$ and $m_\tau = 1784 MeV$ which are in good
agreement with experimental data.

The calculation for the electron EDM is similar to the
neutron EDM.
For case a) we find that the one loop contributions are small ($< 10^{29}
ecm$) with $\sigma' = 80^0$. However the two loop contribution due to Bar-Zee
mechanism\cite{bz,edm}
can be as large as $10^{-27}$ ecm, for $m_t < 200 GeV$.
For case b), we find that the one loop and two loop contributions are small
 ($< 10^{-33}ecm$).

\section
{Conclusions}

We have studied in detail some effects due to two different CP violating
mechanisms in the $S_3\times Z_3$ model. Both mechanisms discussed in this
paper can explain the observed CP violation in the neutral K system.
CP violation in the neutral K system
and the mass difference in the neutral B system constrain the neutral Higgs
boson masses to be in the multi TeV region.  In the previous discussions we
have chosen a particular set of
parameters. The detailed predictions will depend on all the parameters, but the
general features will remain to be the same. We have checked the predictions
using another set of parameters, we find the changes are not significant except
that the electron EDM for case b) can reach $10^{-28}$ ecm. The predictions
presented here represent the typical values for the observables.
We summarize our results for
$\epsilon'/\epsilon$, CP violation in B decays and the neutron and electron
EDMs in Table 1.
It is clear from Table 1, that the predictions for the observables considered
are very different in the MSM and in multi-Higgs doublet models. Future
experiments will be able to rule out some
models.

\vspace{5mm}
\begin{tabular}{|c|c|c|c|}
\hline
Observable&MSM&Case a)&Case b)\\ \hline
$\bar\epsilon$& Input&Input&Input\\ \hline
$\epsilon'/\epsilon$& $10^{-4} \sim 10^{-3}$& $\sim10^{-5}$&$\sim10^{-3}$\\
\hline
B Decay& $\alpha +\beta+\gamma = 180^0$&
$\alpha +\beta+\gamma \neq 180^0$&$\alpha +\beta+\gamma \neq 180^0$\\
Asymmetry&&$sin2\gamma \approx -sin 2\beta \approx 0.2\sim 0.4$&$sin2\gamma
\approx -sin 2\beta \approx 0.25$\\
&&$sin\gamma < 0.1$& $sin\gamma \approx 0$\\ \hline
$D_n$(ecm)& $10^{-31}\sim 10^{-33}$& can reach $10^{-25}$&can reach
$10^{-25}$\\ \hline
$D_e$(ecm)& $<10^{-36}$ & $< 10^{-27}$&$< 10^{-28}$\\ \hline
\end{tabular}

This research was supported by the Department of Energy Grant No.
DE-FG06-85ER40224. One of the authors (NGD) would like to acknowledge the local
hospitality at the Institute of Mathematical Science.

\end{document}